\documentclass[amsmat,amssymb,amsfonts,aps,prb,twocolumn,showpacs]{revtex4}

\usepackage{graphicx}
\usepackage{dcolumn}
\usepackage{bm}
\begin{document}

\title{Coulomb blockade in electron transport through a C$_{60}$ 
molecule from first principles}
\author{J. J. Palacios}
\affiliation{Departamento de F\'{\i}sica Aplicada and Instituto Universitario de Materiales de Alicante (IUMA), Universidad de Alicante, San Vicente del Raspeig, Alicante 03690, Spain.}

\date{\today}

\begin{abstract}
We present results of spin-unrestricted first-principles quantum transport for a gated C$_{60}$ molecule weakly contacted to Al electrodes, making emphasis on the role played by the electronic localization and the spin degree of freedom. As expected, the conductance presents a series of peaks as a function of a gate voltage, demonstrating that transport in the Coulomb blockade regime can be properly treated within a first-principles scheme. A well-known manifestation of the interplay between Coulomb interaction and the spin degree of freedom in atoms and molecules, the Hund 's rule, determines the sequence of conductance peaks.  
\end{abstract}

\maketitle

\section{Introduction}
Since molecular electronics\cite{Aviram:cpl:74} is the focus of attention on the part of investigators in search of the ultimate device miniaturization, many attempts have been made to contact molecules to metallic electrodes. With the appropriate end groups these molecules can bind to the electrodes and form fairly stable molecular bridges as initially shown by Reed et al.\cite{Reed:science:97}. Since then, a combination of chemical functionalization and nanofabrication techniques are being pursued with different degrees of success\cite{Nitzan:science:03}. Similarly, the road to describe and understand theoretically how these contacted molecules carry current is plagued with technical and conceptual difficulties\cite{Palacios:ctcc:05}. One of these difficulties stems from the use of density functional theory (DFT), on which most of the codes developed for the calculation of "first-principles" transport are currently based\cite{Palacios:prb:01,Taylor:prb:01:mar,Brandbyge:prb:02,diVentra:prl:01,Thygesen:prl:05,Sanvito:05}. The notorious discrepancies between experiments and theory are being attributed by some groups to the failure of  standard DFT in describing the behaviour of the quasiparticles at the Fermi level when charge density variations are pronounced\cite{Burke:05,Reimers:anwas:03,Sai:prl:05}. 

While failures attributed to the use of DFT might be critical on cases where strong charge localization occurs and the basis for this criticism is worth pursuing further, it is also imperative to understand the role played by all the others factors at play. One of the factors that has not been fully (not even partially) explored is the spin. It is quite obvious that
magnetism is not a lesser factor when it comes to device functionality. While a large amount of experimental work has been done in large area magnetic junctions and magnetic multilayers due to the huge technological impact of these systems\cite{Wolf:science:01}, a deep theoretical understanding of electron transport in these systems is still lacking. Magnetic nanocontacts, for instance, exhibit a very rich and complex behavior which, at this moment,  is far from being understood\cite{Garcia:prl:99,Viret:prb:02,Jacob:04}. Some nonmagnetic metallic nanocontacts are also expected to exhibit weak magnetism due to the low coordination of the atoms at the breaking point\cite{Delin:prl:04}. Magnetic molecules are also the subject of several experimental studies from the  transport point of view\cite{Park:nature:02}. Furthermore, even if all the elements composing a molecular bridge, namely, the electrodes and the molecule, are not intrinsically magnetic, charge density localization due to poor coupling of the molecule to the electrodes can help localize a spin density at the molecule. This last scenario is our focus in this work. 

\begin{figure}
\centering
\includegraphics[scale=0.25]{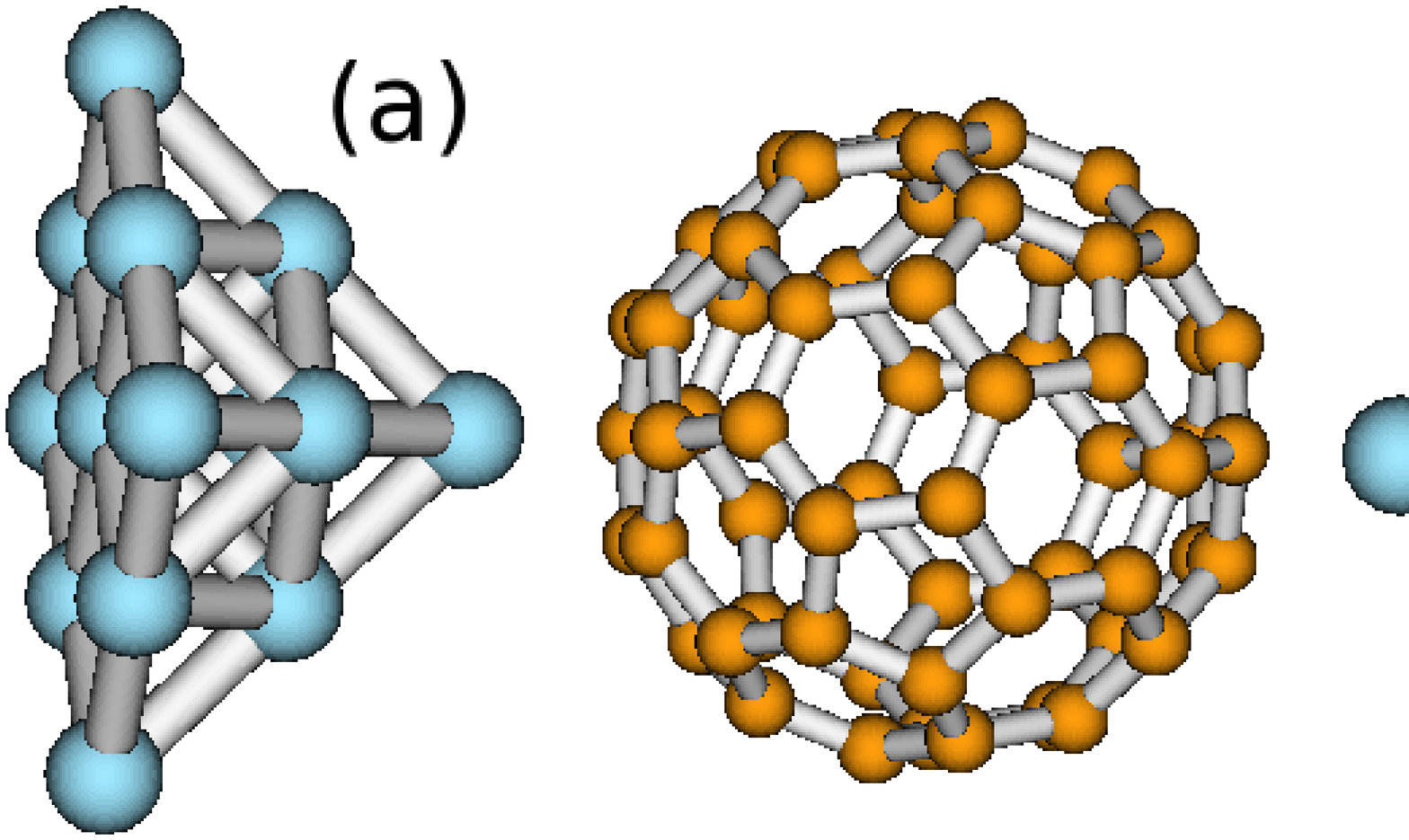}
\includegraphics[scale=0.25]{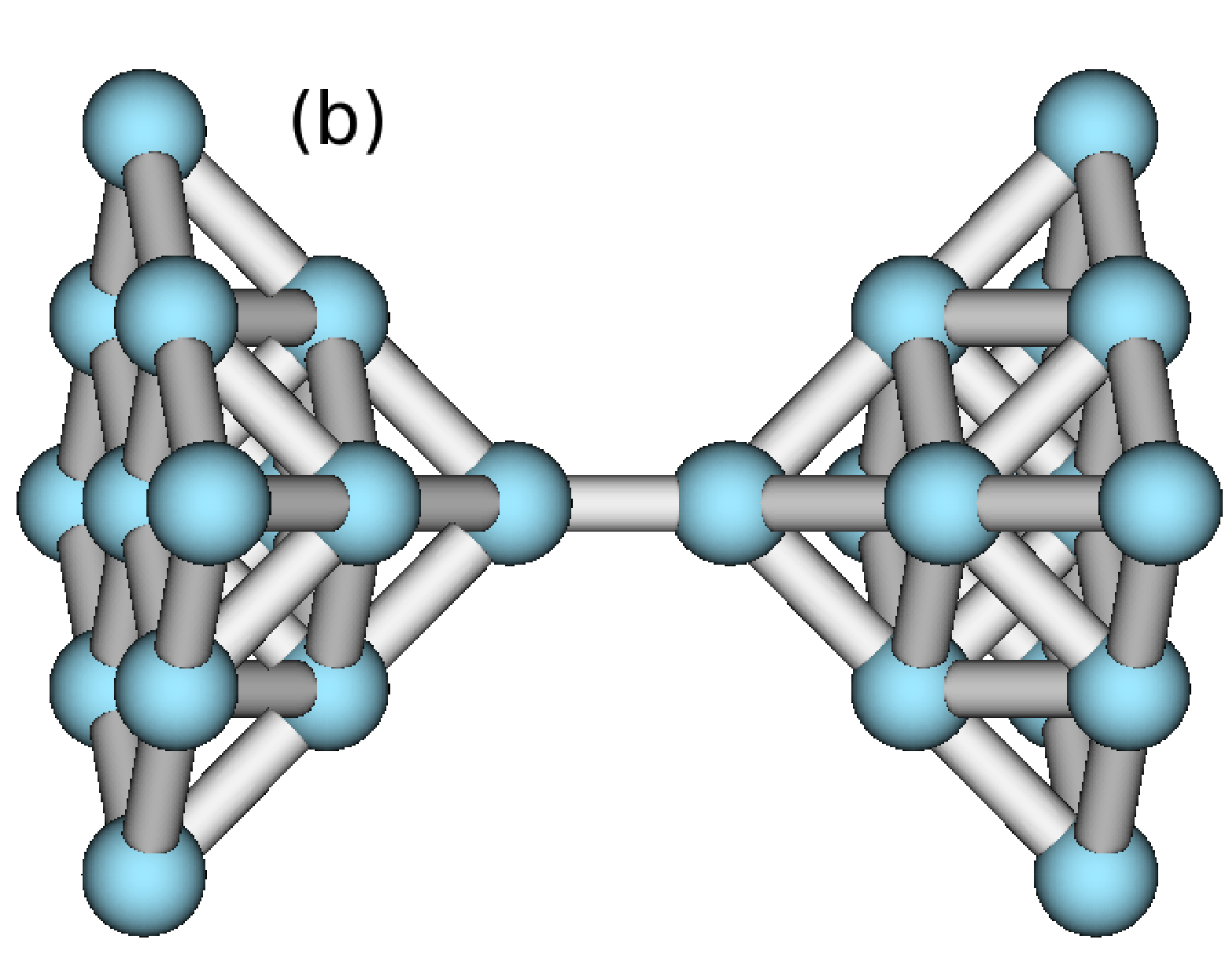}
\caption{(a) A molecular bridge formed by  a C$_{60}$
molecule weakly attached to Al electrodes with pyramidal form. Coulomb blockade is expected to occur due to the formation of quasi-localized molecular orbitals. (b) An Al nanocontact where electronic localization is not expected.}
\label{bridge}
\end{figure}

To date, most first-principles codes for  electronic transport do not take into account the spin degree of freedom since this adds an important degree of difficulty to the well-known computational complexity of this type of calculations (for exceptions see Refs. [\onlinecite{Sanvito:05,Bagrets:prb:04,Smogunov:ss:02}]). Here, we present results obtained from the spin-unrestricted version of our GAUSSIAN03-based\cite{Gaussian:03} \textit{ab initio} code for quantum transport named ALACANT (ALicante \textit{Ab initio} Computation Applied to NanoTransport)\cite{ALACANT:05}. For illustration purposes, we consider here a prototypical molecular bridge: A C$_{60}$ molecule (weakly) contacted to Al electrodes and 
in the presence of a nearby metallic gate [see Fig. \ref{bridge}(a)]. 
Transport through a C$_{60}$ molecule has been previously studied theoretically for Al and Au electrodes with first-principles methods, but without considering separate spin transport channels\cite{Palacios:nanotech:01,Palacios:prb:01,Taylor:prb:01:mar,Sergueev:03}. We explore here the importance of the spin channel separation as well as that of recalculating  the conductance as the Fermi energy is changed by a gate voltage $V_g$.  If the C$_{60}$ molecule is weakly contacted to the electrodes, the charge at the molecule gets strongly localized and transport occurs in the Coulomb blockade regime. 
Despite of bulk Al being nonmagnetic, spin is relevant for transport through the weakly bonded C$_{60}$ molecule. The ground state electronic configuration of an isolated C$_{60}$ molecule is nonmagnetic, with a three-fold degenerate lowest unoccupied molecular orbital (LUMO), $t_u$,  and a five-fold degenerate highest occupied molecular orbital (HOMO), $h_u$. Charge transfer from the electrodes or changes in the Fermi energy induced by nearby gates can change the charge state of the molecule and, therefore, the molecular electronic configuration can become magnetic. In general terms, transport through weakly contacted molecules presents zero-bias conductance peaks as a function of  $V_g$\cite{Park:nature:02,Liang:prl:02,Champagne:nanolett:05,Moriyama:prl:05}. The molecule  gets filled or emptied one electron at a time by the action of this voltage, opening and closing shells (almost) succesively. Thereby the necessity of considering the spin degree of freedom in these situations. In addition to that, we find that Hund 's rule determines the sequence of conductance peaks, something that has been observed in poorly contacted nanotubes by several groups\cite{Liang:prl:02,Moriyama:prl:05}. To our knowledge, this is the first time that transport in the Coulomb blockade regime is fully analyzed from first principles.

\section{THEORETICAL BACKGROUND}
\subsection{Green's function formalism for open-shell systems}
The design and fabrication of electronic devices at the molecular and atomic scale has posed new challenges to theorists. The basics to calculate the zero-bias, zero-temperature conductance, $\mathcal{G}$,  in a molecular bridge or metallic nanocontact were  established by Landauer long before the concept of nanoelectronics was commonplace.  In Landauer's formalism $\mathcal{G}$ is proportional to the quantum mechanical transmission probability for the electrons at the Fermi energy, $E_{\rm F}$, to cross the molecular bridge\cite{Datta:book:95}:
\begin{equation}
\mathcal G=\frac{e^2}{h} [T_\uparrow (E_{\rm F})+T_\downarrow (E_{\rm F})]. 
\label{g}
\end{equation}
In this expression the contributions from spin up ($\uparrow$) and spin down ($\downarrow$) channels have been explicitly separated while the contribution from all the orbital channels has been condensed in $T$. For simplicity we neglect spin mixing due to spin-orbit scattering and exclude the possibilty of noncollinear spin densities, i.e., $S_{\rm z}$ is a good quantum number. It is well-known that the detailed atomic, electronic, and magnetic structure at a  nanocontact\cite{Jacob:04} is important and, in order to achieve a quantitative level of agreement with experiments, one has to rely on first-principles or \textit{ab initio} calculations, typically at the DFT level. For molecular bridges like the one in Fig. \ref{bridge}(a) this necessity becomes imperative due to the impossibility of predicting (i) the broadening of these orbitals due to the coupling with the electrodes and, most importantly, (ii) the positioning of the Fermi level with respect to the HOMO and LUMO of the molecule. Two different approaches to this problem can be found in the literature: One based on calculating scattering wave functions\cite{Lang:prb:95,Hirose:prb:95,diVentra:prl:01}
and the other based on Green function techniques\cite{Palacios:prb:01,Palacios:prb:02,Yaliraki:jcp:99,Xue:jcp:01,Taylor:prb:01:mar,Thygesen:prl:05,Ke:prb:04,Jelinek:prb:03,Heurich:prl:02}, sometimes combined with the Keldysh formalism\cite{Louis:prb:03,Taylor:prb:01:dec,Brandbyge:prb:02}. The basics of our approach, which we extend below to include the spin degree of freedom, have been presented in previous publications\cite{Palacios:prb:01,Palacios:prb:02,Louis:prb:03}. 

The main advantage of the numerical implementation that we have developed is that relies on a standard quantum chemistry code such as GAUSSIAN03. This code is a versatile tool to perform first-principles or \textit{ab initio} calculations of molecules or clusters that incorporates the major advancements in the field in terms of functionals, basis sets, pseudopotentials, etc.. The way to proceed is as follows: A standard self-consistent field electronic structure calculation of the molecule that includes a significant part of the electrodes [like that shown in Fig. \ref{bridge}(a)] is performed. This calculation is usually performed at a DFT level in any of its multiple approximations. The use of configuration interaction or multiple Slater determinant methods  has been recently 
proposed to study regimes of transport beyond the scope of DFT such as the Kondo effect\cite{Davidovich:prb:02} or simply to improve existing DFT results\cite{Ferretti:prl:05}, but these approaches limit the calculations to very simple systems and present some conceptual difficulties not fully resolved. 

As far as transport is concerned, the self-consistent hamiltonian  $\mathbf H_{\uparrow(\downarrow)}$ (or Fock matrix $\mathbf F_{\uparrow(\downarrow)}$) of the central cluster contains the relevant information. The retarded (+) and advanced (-) Green's functions  associated with the  Fock matrices are defined in a standard manner for both spin up and spin down species:
\begin{equation}
\mathbf{G}_{\uparrow(\downarrow)}^{(\pm)}= 
\left [(E\pm i\delta)\mathbf{1}- \mathbf{F}_{\uparrow(\downarrow)}\right ]^{-1} 
\end{equation}
In this expression $\mathbf{1}$ is the identity matrix and $\delta$ is an infinitesimal quantity (in practice we set it to $10^{-10}$ eV).
One of the many advantages in the use of Green functions is that one can 
incorporate the rest of the infinite electrodes in the calculation
in a very elegant and simple way:
\begin{equation}
 \mathbf{G}_{\uparrow(\downarrow)}^{(\pm)}(E)=
\left [(E\mathbf{1}- \mathbf{F}_{\uparrow(\downarrow)} - \mathbf{\Sigma}_{\uparrow(\downarrow)}^{(\pm)}(E)
\right ]^{-1}, 
\end{equation}
\noindent  where
\begin{equation}
\mathbf{\Sigma}_{\uparrow(\downarrow)}^{(\pm)}(E)=\mathbf{\Sigma}_{\rm R\uparrow(\downarrow)}^{(\pm)}(E) +\mathbf{\Sigma}_{\rm L\uparrow(\downarrow)}^{(\pm)}(E), 
\end{equation}
\noindent and $\mathbf{\Sigma}_{\rm R_{\uparrow(\downarrow)}}(E)$[$\mathbf{\Sigma}_{\rm L_{\uparrow(\downarrow)}}(E)$]
denotes a self-energy matrix that represents the spin up(down) hamiltonian of the right(left) semi-infinite electrode that has not been explicitly included in the calculation. With non-orthogonal basis sets, which are the ones implemented in GAUSSIAN03, it has become custommary
to use the following expression for the Green function:
\begin{equation}
\mathbf{G}_{\uparrow(\downarrow)}^{(\pm)}(E)= \left [(E\mathbf{S}- \mathbf{F}_{\uparrow(\downarrow)} - \mathbf{\Sigma}_{\uparrow(\downarrow)}^{(\pm)}(E)
\right ]^{-1}
\label{gno}
\end{equation}
where the overlap matrix $\mathbf S$ essentially substitutes $\mathbf 1$. The self-energy matrices can only be explicitly calculated in ideal situations. We describe the bulk electrode with a parametrized tight-binding Bethe lattice model with the coordination number and effective parameters appropriate for the type of electrodes. The advantage of a Bethe lattice resides in that it reproduces fairly well the bulk density of states of most commonly used metallic electrodes and, at the same time, it does not represent a specific atomic arrangement which is also desirable; for electrodes are polycrystalline.
The expression for the Green function in a non-orthogonal basis set presented in Eq. \ref{gno} is widely accepted in the literature, although it does not strictly correspond to the representation of the Green function operator in a non-orthogonal set\cite{Louis:prb:03}. It is, however, convenient since the density matrix $\mathbf P_{\uparrow(\downarrow)}$ can be obtained from it in the standard way:
\begin{equation}
\mathbf{P}_{\uparrow(\downarrow)}=-\frac{1}{\pi}\int_{-\infty}^{E_{\rm F}}{\rm Im}
\left[\mathbf{G}_{\uparrow(\downarrow)}^{(+)}(E)\right ]\;{\rm d}E.
\label{eqn:nab}
\end{equation}
One can now calculate  the total electron charge of the cluster:
\begin{equation}
N={\rm Tr}[ (\mathbf{P}_\uparrow +\mathbf{P}_{\downarrow}) \mathbf{S}].
\label{charge}
\end{equation}
In Eq. \ref{charge} the trace runs over all the atomic orbitals or localized functions 
composing the basis set.  In Eq. \ref{eqn:nab} $E_{\rm F}$ is determined by fixing the total charge in the cluster, $Q=N+N_{\rm ion}$, where $N_{\rm ion}$ is the fixed ion charge. In practice $Q$ needs to be distributed between both spin species to comply with the thermodynamic constrain of a unique Fermi level for both spin channels. We force GAUSSIAN03 to evaluate $F_{\uparrow}$ and $F_\downarrow$ using the density matrices obtained from Eq. \ref{eqn:nab} and repeat the process until convergence is achieved. 
The spin-dependent transmission probabilities that appear in Eq. \ref{g} can be calculated through the well-known expression:
\begin{equation}
 T(E) =
{\rm Tr}[\mathbf{\Gamma}_{\rm L}(E) \mathbf{G}^{(+)}(E)\mathbf{\Gamma}_{\rm R}(E) \mathbf{G}^{(-)}(E)],
\label{t} 
\end{equation}
where, again,  the trace runs over all the orbitals and $\mathbf{\Gamma}_{\rm L,R}(E)$ are  twice the imaginary part of the self-energy matrices. Finally, in order to single out the contribution of individual orbital channels to the current, one can diagonalize the  matrix product in Eq. \ref{t}. While the size of the resulting product matrix in brackets  can  be as large as desired, the number of eigenvalues with a  significant contribution, i.e., the number of conducting channels, will be typically much smaller. 

\begin{figure}
\centering
\includegraphics[scale=0.3]{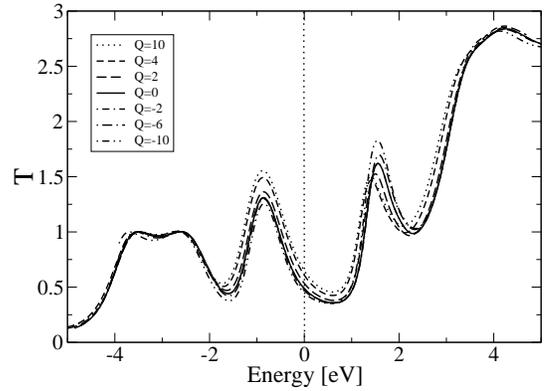}
\caption{Sequence of conductance curves for different values of the total charge $Q$ in the nanocontact shown in Fig.\ \ref{bridge}(b). The curves have been shifted along the energy axis to make them coincide as much as possible.\label{nanocontact}}
\end{figure}

\subsection{Conductance as a function of gate voltage}
Although Eq. \ref{t} gives the transmission as a function of energy, only the value of $T$ at the Fermi energy is, in general, meaningful (within the limitations inherent to DFT\cite{Sai:prl:05,Burke:05}).  For instance, if a metal plate is placed in the vicinity of the molecular bridge and a gate voltage $V_g$ is applied to it, the total charge of the molecular bridge, $Q$, changes in response to this voltage, as in a classical capacitor. Noticing that the gates are typically much larger than the molecule, changing the voltage changes the net charge of the overall molecule+electrodes system and not only of the molecule\cite{Ke:prb:05}. The relation between the magnitude of $V_g$ and the total charge accumulated in the bridge is determined by the capacitance, and this is determined, in turn, by the geometry of the whole molecular bridge-metallic gate set-up. Having knowledge of this capacitance can be of interest, but it is not relevant for our purpose here. Therefore, we simply use the net charge of the cluster as the independent variable. The whole curve $T(E)$ thus needs to be recalculated for each new value of the total charge, which, in turn, determines a new Fermi energy. In general, 
$T(E)$ for $E\neq E_{\rm F}$ does not represent a true measurable quantity. Naively, though, one can be tempted to describe the  effects of this charging as a rigid shift of the density of states (DOS) and, consequently, 
\begin{equation}
 T(E,V_g) = T(E-\gamma eV_g),
 \label{shift}
\end{equation}
where $\gamma$ represents a scaling factor associated with the specific geometry. This way of effectively accounting for the gate voltage  is valid for bridges with a very smooth DOS
such as that in purely metallic nanocontacts as the one shown in Fig. \ref{bridge}(b).
Figure \ref{nanocontact} shows $T(E)$ for various values of $Q$. In this case the results essentially confirm the naive behavior expected from Eq. (\ref{shift}). On the other hand, if localized or quasi-localized states (e.g., molecular states) cross the Fermi energy in performing the energy shift, Eq. (\ref{shift}) no longer can be applied. Coulomb repulsion pushes up or down all the empty states, depending on whether a molecular orbital fills up or empties.  This is typically the case when the molecule is loosely attached to the electrodes. Charging effects on the molecule drastically change $T(E)$  as the Fermi energy varies even by small amounts. The next section illustrate what, in general, can be stated as  
\begin{equation}
T(E,V_g) \neq T(E-\gamma eV_g).
\end{equation}

\section{COULOMB BLOCKADE in transport through a C$_{60}$ molecule}
\begin{figure}
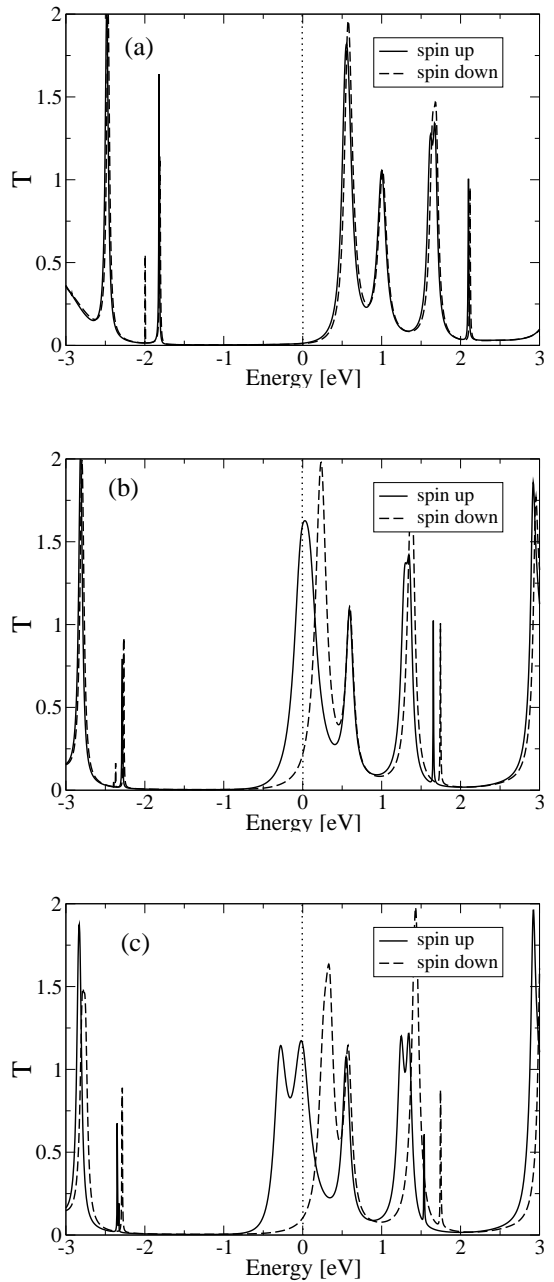

   \centering
   \includegraphics[scale=0.3]{T-0.eps}\\
   \vspace{0.75cm}
   \includegraphics[scale=0.3]{T-6.eps}\\
   \vspace{0.75cm}
   \includegraphics[scale=0.3]{T-7.eps}
   \caption{Sequence of transmission curves as a function of the overall charge in the whole
   cluster. The electronic charge increases from top to bottom, starting from a positively charged cluster: (a) $Q=10$, (b) $Q=4$, and (c) $Q=3$.}
   \label{fig1:T}
\end{figure}

\begin{figure}
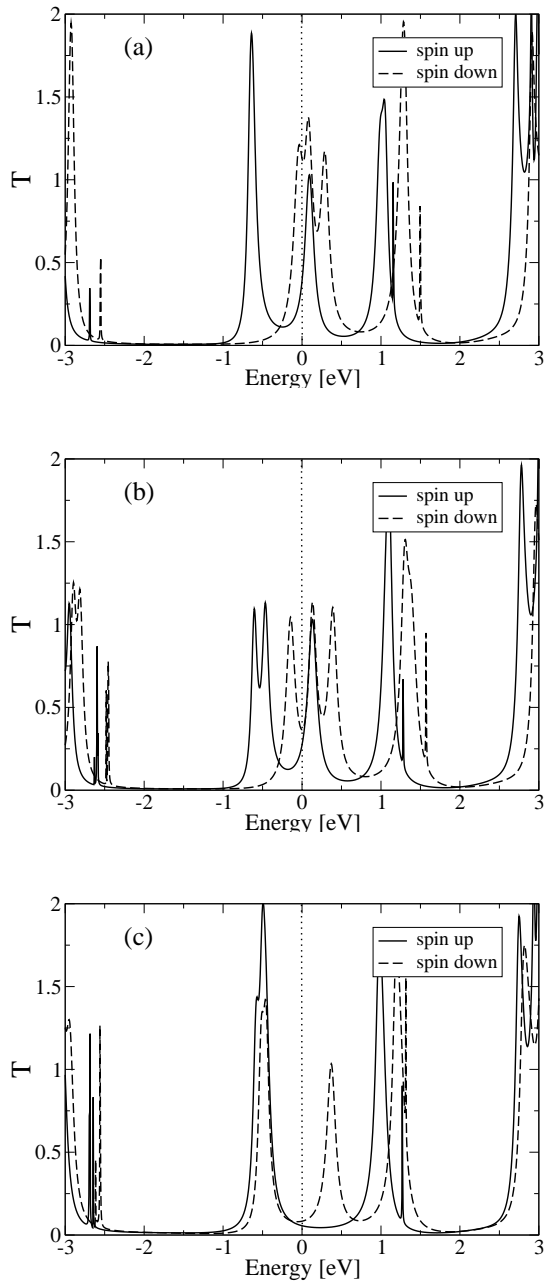

   \includegraphics[scale=0.3]{T-13.eps}\\
   \vspace{0.75cm}
   \includegraphics[scale=0.3]{T-14.eps}\\
   \vspace{0.75cm}
   \includegraphics[scale=0.3]{T-20.eps}
   \caption{Same as in Fig. \ref{fig1:T} for further increase of the total electronic charge: (a) $Q=-3$, (b) $Q=-4$, and (c) $Q=-10$.}
   \label{fig2:T}
\end{figure}

Figures \ref{fig1:T} and \ref{fig2:T} show the zero-temperature transmission $T$ as a function of the scattering energy for the molecular bridge shown in Fig. \ref{bridge}(a). We have chosen a pyramidal model for the electrodes in order to have a weak contact to the molecule. Similar atomic structures for the electrodes close to the contact with the molecule are, nevertheless, expected to form in most break-junction experiments\cite{Rego:prb:03}. Our present calculations are based on DFT theory, but making use of the B3LYP hybrid functional\cite{Gaussian:03}. The basis sets and pseudopotentials are those of Christiansen et al.\cite{Pacios:jcp:85,Hurley:jcp:86}. This combination of basis set and functional has been proved to give good results at a reasonable computational cost in a variety of situations\cite{Palacios:prb:01,Palacios:prb:02,Louis:prb:03,Garcia:prb:04}. The accuracy of our calculations can be systematically improved by resorting to larger basis sets, but this is not  essential for our discussion here. The use of the B3LYP functional provides us, in addition, with a semi-phenomenological way of dealing with the well-known problems of local density approximations to the true functional. The exact non-local Hartree-Fock potential is included in this functional, partially taking care of the self-interaction problem and inherent delocalization of charge typical in local or quasi-local
approximations. This way the value of the transmission off resonance and the shape of the Coulomb blockade peaks is expected to approach to the true ones compared to the results obtained with LDA or even GGA.

The six panels in Figs. \ref{fig1:T} and \ref{fig2:T} present $T$ for both spin up (solid lines) and spin down (dashed lines) channels for increasing Fermi energy or increasing charge (top to bottom). It has been shown that a C$_{60}$ molecule contacted to Al electrodes gets charge transferred from the electrodes in equilibrium and without the effect of any gate voltages\cite{Palacios:prb:01}.  This is a purely chemical effect. In panel (a) of Fig. \ref{fig1:T} the Fermi level has been shifted down with respect to the neutrality point, corresponding to a slightly discharged molecular bridge. There the Fermi level lies in the HOMO-LUMO gap of the molecule so that the bare molecule remains neutral and nonmagnetic. In this panel the transmission reveals two degenerate spin channels with a series of peaks corresponding to molecular states broadened by the coupling to the electrodes. The conductance peaks below the Fermi energy reflect the DOS corresponding to occupied states and those above to empty ones. From now on we focus on the empty ones: The three spin-degenerate $t_u$ molecular states. For the isolated and neutral molecule these states are fully degenerate. In panel (a) of Fig. \ref{fig1:T}we see that the coupling to the electrodes not only broadens these levels, but also lifts partially the degeneracy of them. The symmetry of these states combined with the chosen bonding geometry to the metal contacts  splits these three levels in 2+1. This reflects in a transmission peak reaching a value of two around 0.6 eV and one reaching a value of one around 1.0 eV for both spin channels. More molecular states, those termed $t_g$, also reflect in the transmission at higher energies. 

As the Fermi energy is increased and the doubly-degenerate states start to fill up [see Fig. \ref{fig1:T} (b)], the spin degeneracy also begins to be lifted. This broken spin symmetry is a consequence of our mean field level description. Spin symmetry preserving calculations, which may give rise to Kondo physics at low temperatures, are out of the scope of our present methodology. It must be understood in what follows that the shape and height of the peaks would be changed by the Fermi distribution function above the Kondo temperature\cite{Hewson:book}, which can be tuned down to zero by decreasing the coupling of the molecule to the electrodes. (We have selected a case with a relatively strong coupling for clarity). On further increasing the net charge [panel (c)] and, consequently, the Fermi energy, the spin degeneracy is fully removed and the orbital degeneracy of the doubly-degenerate $t_u$ orbitals is also lifted as one of them charges up. Again, this broken symmetry is a consequence of the mean-field approximation. Notice that the spin and orbital symmetry breaking brings the maximum of the conductance peaks down to the quantum of conductance $e^2/h$, as expected for resonant tunneling through an orbital symmetrically coupled to both electrodes (in the absence of the Kondo effect). Notice, finally, how the second molecular orbital that gets filled has the same spin as the first one. This is what is expected from Hund 's rule and agrees with what various calculations in the literature have predicted for C$_{60}$ negative ions\cite{Martin:prb:93}. As the Fermi energy increases further, the doubly-degenerate spin-down orbitals get their degeneracy removed by the charging of one of them while the Coulomb interaction pushes the other up in energy [see Fig. \ref{fig2:T}(a)]. From here on [Figs. \ref{fig2:T}(b) and (c)], the remaining spin up orbital and the other two empty spin down orbitals get filled one at a time. We would like to conclude our analysis by noting that the specific order in which the $t_u$ orbitals get filled depends very much on the bonding geometry to the electrodes and the type of electrodes. Also the reliability of the B3LYP approximation to DFT in describing the transmission at the Fermi energy in the Coulomb blockade regime needs to be tested more thouroghly. More sistematic work on Coulomb blockade in molecular bridges is obviously required.

\section{CONCLUSIONS}
We have studied a prototypical example of a molecular bridge: a C$_{60}$ molecule contacted by Al electrodes. It has been shown that spin-unrestricted \textit{ab initio} transport calculations in weakly contacted molecular bridges, where localization of charge is expected to occur, can reproduce the main features of the Columb blockade phenomenon. In regards to this, it has been shown the necessity of recalculating transmission curves as the Fermi energy changes in systems that exhibit charge localization. Transmission curves calculated at different Fermi energies can be completely different due to charging effects. In addition, Hund 's rule has been shown to play an important role in the transport characteristics.

\section{Acknowledgments}
We acknowledge dicussions with J. Fern\'andez-Rossier and David Jacob. Financial support from Grant No. MAT2002-04429-C03 (MCyT) and from the University of Alicante is also acknowledged.



\end{document}